\documentclass{pasj00}

\begin{document}
\SetRunningHead{T. Kato and Y. Osaki}{GALEX J194419.33$+$491257.0}

\Received{201X/XX/XX}
\Accepted{201X/XX/XX}

\title{GALEX J194419.33$+$491257.0: An Unusually Active
SU UMa-Type Dwarf Nova with a Very Short Orbital Period
in the Kepler Data}

\author{Taichi \textsc{Kato}}
\affil{Department of Astronomy, Kyoto University,
       Sakyo-ku, Kyoto 606-8502}
\email{tkato@kusastro.kyoto-u.ac.jp}

\and

\author{Yoji \textsc{Osaki}}
\affil{Department of Astronomy, School of Science, University of Tokyo,
Hongo, Tokyo 113-0033}
\email{osaki@ruby.ocn.ne.jp}



\maketitle

\begin{abstract}
   We studied the background dwarf nova of KIC 11412044
in the Kepler public data and identified it with
GALEX J194419.33$+$491257.0.  This object turned out
to be a very active SU UMa-type dwarf nova having
a mean supercycle of about 150~d and frequent normal outbursts
having intervals of 4--10~d.  The object showed strong
persistent signal of the orbital variation with a period
of 0.0528164(4)~d (76.06~min) and superhumps with a typical period
of 0.0548~d during superoutbursts.  Most of the superoutbursts
were accompanied by a precursor outburst.  All these features
are unusual for this very short orbital period.
We succeeded in detecting the evolving stage of superhumps
(stage A superhumps) and obtained a mass ratio of 0.141(2),
which is unusually high for this orbital period.
We suggest that the unusual outburst properties are
a result of this high mass ratio.  We suspect that this
object is a member of the recently recognized class of
cataclysmic variables (CVs) with a stripped core evolved secondary 
which are evolving toward AM CVn-type CVs. 
The present determination of the mass ratio using stage A
superhumps makes the first case in such systems.
\end{abstract}

\section{Introduction}
   The Kepler mission (\cite{bor10Keplerfirst}; \cite{Kepler}),
which was aimed to detect extrasolar planets, has provided
unprecedentedly sampled data on several cataclysmic variables (CVs).
This satellite also recorded previously unknown CVs
as by-products of the main target stars.  The best documented
example has been the background dwarf nova of KIC 4378554
(\cite{bar12j1939}; \cite{kat13j1939v585lyrv516lyr}).
In addition to this object, the group Planet Hunters
\citep{fis12PlanetHunters} detected several candidate
background CVs.\footnote{
  $<$http://talk.planethunters.org/objects/APH51255246/\\
discussions/DPH101e5xe$>$.
}  

   We studied one of these background dwarf novae,
the one in the field of KIC 11412044 (hereafter J1944).
This object was discovered by the Planet Hunters group as a background 
SU UMa-type dwarf nova of KIC 11412044, in which superoutbursts and 
frequent normal outbursts were recognized. \footnote{
  $<$http://keplerlightcurves.blogspot.jp/2012/07/\\
  dwarf-novae-candidates-at-planet.html$>$.
}  
Since it was bright enough and it was frequently
included in the aperture mask of KIC 11412044, the outburst
behavior can be immediately recognized in Kepler {\tt SAP\_FLUX}
light curve of KIC 11412044.

\section{Data Analysis}

\begin{figure*}
  \begin{center}
    \FigureFile(160mm,230mm){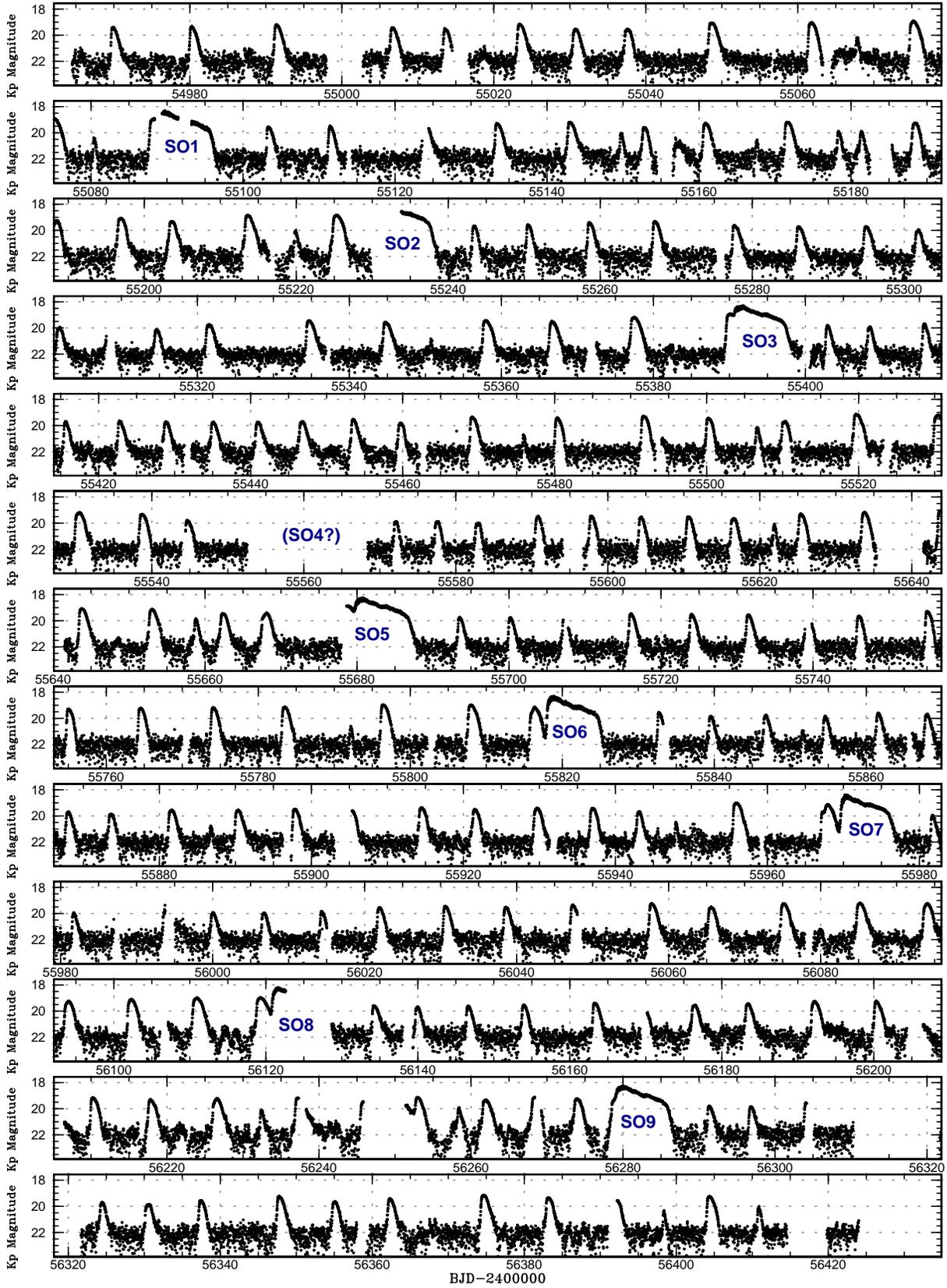}
  \end{center}
  \caption{The Kepler LC light curve of J1944.
  The superoutbursts are marked with labels.
  }
  \label{fig:j1944lc}
\end{figure*}

\begin{figure*}
  {\centering
    \FigureFile(160mm,120mm){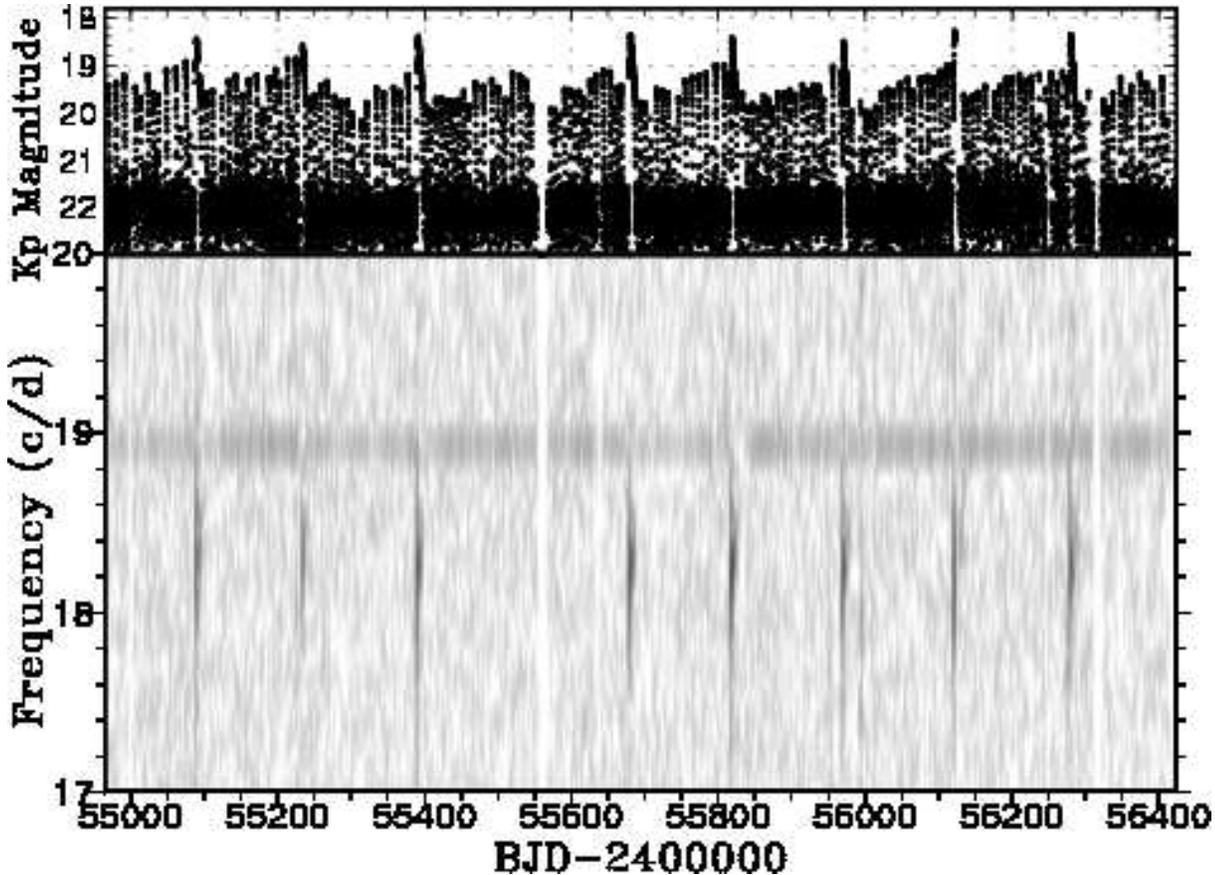}
  }
  \caption{Two-dimensional Fourier power spectrum of the Kepler
  LC light curve of J1944.
  (Upper:) Light curve; the Kepler data were binned to 0.08~d.
  (Lower:) Fourier power spectrum. The width of 
  the sliding window and the time step used are 10~d and 1~d,
  respectively.  The signal at frequency 18.93 cycle d$^{-1}$ is constantly
  seen, and we interpret it as the orbital period.
  The broad signals around 18.1--18.5 cycle d$^{-1}$ are superhumps.
  }
  \label{fig:j1944spec2d}
\end{figure*}

\begin{figure}
  {\centering
    \FigureFile(88mm,88mm){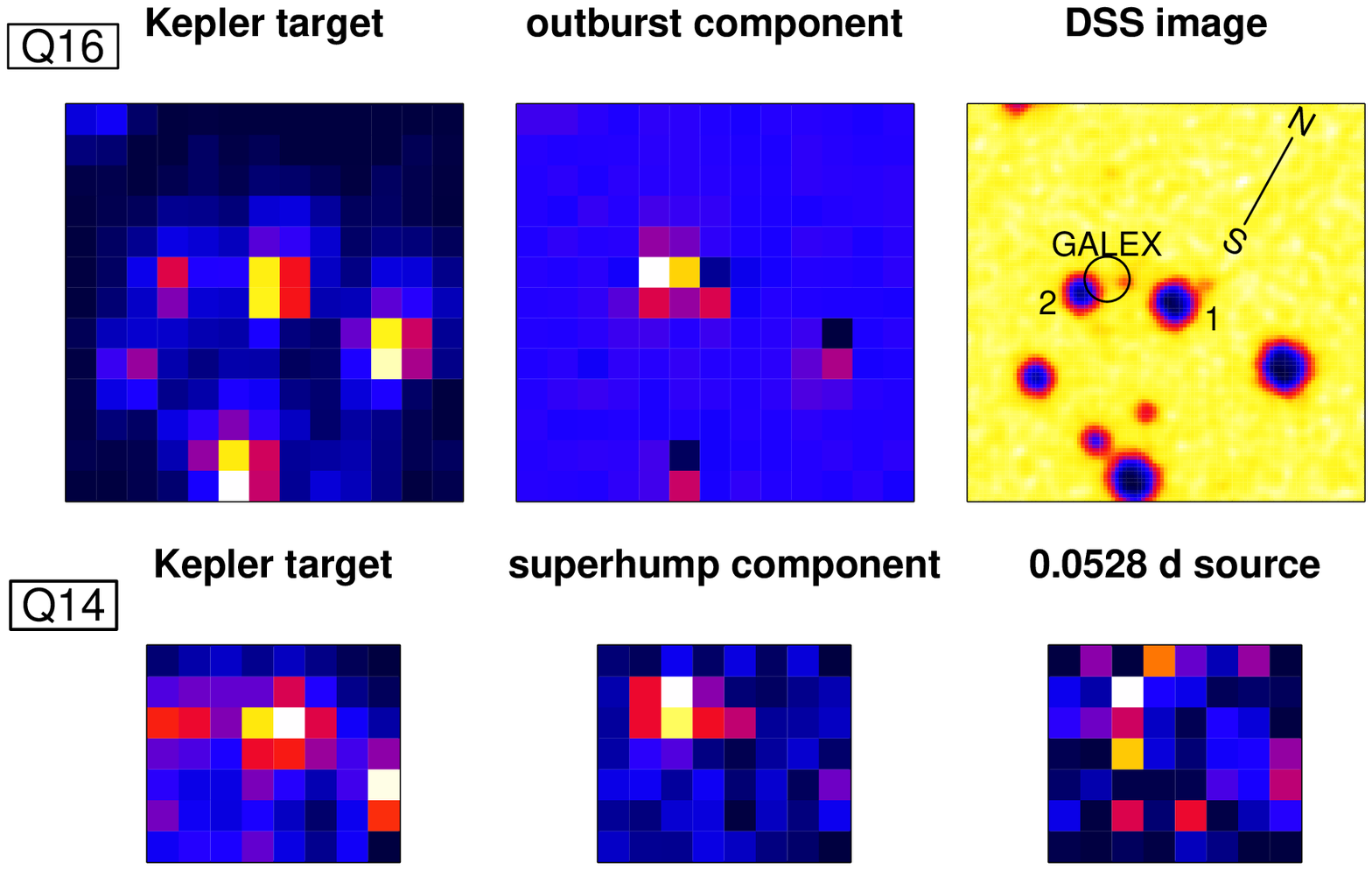}
  }
  \caption{
     (Upper panels): Q16 Kepler target pixel images (13$\times$13 pixels).
     Brighter colors represent stronger signals except for
     the digitized sky survey (DSS) image.  From left to right,
     flux of Kepler target pixels, outbursting component
     and the corresponding DSS image (the color level was
     reversed for better visibility).  The main target
     KIC 11412044 (star 1 on the DSS image) is located 
     at the center of the left image.
     There is a fainter star (star 2 on the DSS image)
     three pixels left to KIC 11412044.
     The middle image was created by subtracting the image
     in quiescence from the outburst image.  The outbursting
     component is located two pixels (8 arcsec) left to
     KIC 11412044.  There is a faint star between KIC 11412044
     and star 2 on the DSS2 (blue plate, rotated to match the Kepler
     imaging direction) image (right image).
     This position is in good agreement with the location
     of the GALEX UV source. \\
     (Lower panels): Q14 Kepler target pixel images (8$\times$7 pixels).
     From left to right, flux of Kepler target pixels,
     superhumping component and 0.0528~d component.
     The locations of superhumping and 0.0528~d components
     are in good agreement with the outbursting component,
     i.e. two pixels left to KIC 11412044.
     These quarters were selected because Q16 has the widest
     field of view and Q14 contained a superoutburst.
     The pixel size of Kepler images is 3.98$\times$3.98 arcsec.
     }
  \label{fig:kepmap}
\end{figure}

   We used Kepler public long cadence (LC) data (Q1--Q17)
for analysis.  Since the outbursts were immediately recognizable
in each light curve of the Kepler target pixel images,
we used a custom aperture consisting of 4--6 pixels
showing outbursts as we did in the background dwarf nova
of KIC 4378554 \citep{kat13j1939v585lyrv516lyr}.
We used surrounding pixels to subtract
the background from KIC 11412044.  We further corrected
small long-term baseline variations by subtracting
a locally-weighted polynomial regression (LOWESS: \cite{LOWESS})
and spline functions.
Since the quiescent magnitude is difficult to determine,
we artificially set the level to be 22.0 mag.

\section{Characterization and Identification of Object}

\subsection{Outburst Properties}

   The resultant light curve indicates that this object
is an SU UMa-type dwarf nova with frequent outbursts
(figure \ref{fig:j1944lc}).
There were eight observed superoutbursts,
and from the regular pattern, another superoutburst most likely 
occurred between BJD 2455553 and 2455568 
(a data gap in Q8) and we numbered the superoutburst and supercycle
assuming that there is a superoutburst in this gap.
The intervals between successive superoutbursts (supercycles)
were in a range of 120--160~d.  We determined the mean
supercycle of 147(1)~d.  Most of superoutbursts were
associated with a precursor outburst with a various
degree of separation from the main superoutburst.
The typical duration of the superoutburst is $\sim$8~d
including the precursor part.  This duration is shorter
than those of many other SU UMa-type dwarf novae.

   The number of normal outbursts in one supercycle
ranged from 11 to 21.  The intervals
of normal outbursts were 4--10~d, one of the shortest
known except ER UMa stars \citep{kat99erumareview}.
The amplitudes of normal outbursts increased as 
the supercycle phase progresses.
Some of the normal outbursts were ``failed'', i.e. they
decayed before reaching the full maximum.

\subsection{Frequency Analysis and Source Identification}

   As shown in figure \ref{fig:j1944spec2d}, a two-dimensional 
Fourier analysis (using the Hann window function) of 
the light curve of this object yielded two periods.
There was a signal of a constant frequency (18.93 cycle d$^{-1}$)
with the almost constant strength.  Using all the data
segment, we determined the period to be 0.0528164(4)~d
(18.934 cycle d$^{-1}$).  We refer this signal to ``0.0528~d'' signal. 
   Based on the high stability of the 0.0528~d signal during 
the entire Kepler observations, we identified this period
to be the orbital period ($P_{\rm orb}$) of this object. 
During superoutbursts, there were transient signals of superhumps 
at frequencies around 18.1--18.5 cycle d$^{-1}$ as expected.

   Let us now examine the source position of the background 
dwarf nova in figure \ref{fig:kepmap}.
We checked the pixels which showed the dwarf nova-type
variation.  The peak of signal of dwarf nova-type outbursts
was found two pixels away from the center of KIC 11412044 
(star 1 on the DSS image).
At this location, there is a GALEX \citep{GALEX}
ultraviolet source GALEX J194419.33$+$491257.0
[NUV magnitude 21.3(3)] and
we identified this source as the UV counterpart of this
dwarf nova (figure \ref{fig:kepmap}, Q16), confirming
the suggestion in the Planet Hunters' page.
The superhump component and 0.0528~d component were also
confirmed at the location of this object
(figure \ref{fig:kepmap}, Q14), and we consider that
the 0.0528~d signal indeed comes from this dwarf nova.
This has also been confirmed by the non-detection
of the 0.0528~d signal in the {\tt SAP\_FLUX} of KIC 11412044
when this dwarf nova was outside the aperture of KIC 11412044.

\subsection{Variation of Superhump Period}

\begin{figure}
  \begin{center}
    \FigureFile(88mm,110mm){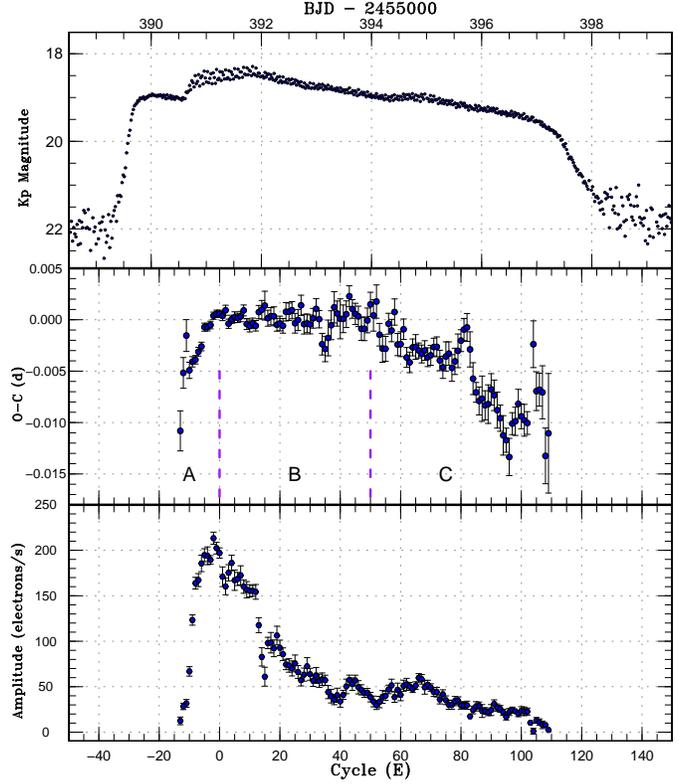}
  \end{center}
  \caption{$O-C$ diagram of SO3 of J1944.
  From top to bottom:
  (1): Kepler LC light curve.
  (2): $O-C$ diagram.
  The figure was drawn against a period of 0.05479~d.
  Stages A--C (cf. \cite{Pdot}) can be recognized.
  Long-period superhump (stage A superhumps)
  with growing superhumps were recorded during
  the late part of the precursor outburst to the maximum 
  of the superoutburst.
  (3): Amplitudes in electrons s$^{-1}$.}
  \label{fig:j1944humpall}
\end{figure}

   Since the superhump period is less than three LC exposures,
it is difficult to determine the times of superhump maxima
by the conventional method.  We employed the Markov-chain 
Monte Carlo (MCMC) modeling used in \citet{kat13j1939v585lyrv516lyr}.
Although we only show the result of SO3
(figure \ref{fig:j1944humpall}), the pattern is similar
in other superoutbursts.  In the $O-C$ diagram, stages A--C
(for an explanation of these stages, see \cite{Pdot})
can be recognized.
Long-period superhumps (stage A superhumps)
with growing superhumps were recorded during
the late part of the precursor outburst to the maximum 
of the superoutburst.  The overall pattern is very similar
to those of other SU UMa-type dwarf novae, including
V1504 Cyg and V344 Lyr (\cite{osa13v344lyrv1504cyg};
\cite{osa14v1504cygv344lyrpaper3}).  The object makes
the fourth case (after V1504 Cyg, V344 Lyr and V516 Lyr)
in the Kepler field in which the growing
superhumps lead smoothly from the precursor to
the main superoutburst and thus gives further support to 
the thermal-tidal instability (TTI) model \citep{osa89suuma}
as the explanation for the superoutburst.

\subsection{System Properties}

   The inferred fractional superhump period excess 
$\varepsilon \equiv P_{\rm SH}/P_{\rm orb}-1$,
where $P_{\rm SH}$ is the superhump period, of $\sim$3.8\%
is, however, unusually large for this $P_{\rm orb}$
(cf. figure 15 in \cite{Pdot}).

\begin{figure}
  {\centering
    \FigureFile(88mm,70mm){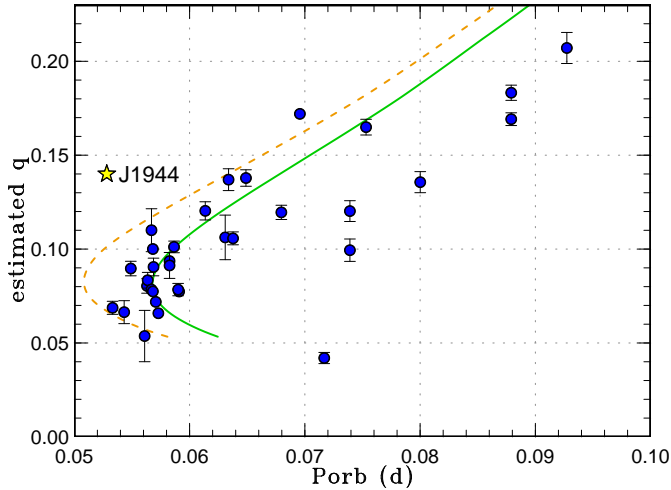}
  }
  \caption{Location of J1944 on the evolutionary track.
  The location of J1944 is plotted (star mark) on figure 5 of
  \citet{kat13qfromstageA}.  The filled circles and
  filled squares represent $q$ values determined using
  stage A superhumps and quiescent eclipses, respectively.
  The dashed and solid curves represent the standard and optimal
  evolutionary tracks in \citet{kni11CVdonor}, respectively.
  }
  \label{fig:evolloc}
\end{figure}

   \citet{kat13qfromstageA} recently proposed that
stage A superhumps can be used to determine 
the mass ratio ($q=M_2/M_1$) and the resultant mass ratios
are as accurate as those obtained from eclipse modeling.
We have succeeded in measuring the period of stage A
superhump during the three superoutbursts: 
0.0555(2)~d (SO3), 0.05546(5)~d (SO6), 0.05552(6)~d (SO7).
The corresponding fractional superhump excesses
in the frequency unit $\varepsilon^* \equiv 1-P_{\rm orb}/P_{\rm SH}$ 
are 4.8\%, 4.77\% and 4.88\%.  These values correspond to the
$q$ value of 0.14, 0.139 and 0.143, respectively.
We therefore adopted $q$=0.141(2).
This mass ratio implies a massive (approximately two times
more massive) secondary for this very short
orbital period comparable to most WZ Sge-type dwarf novae
(figure \ref{fig:evolloc}).
This result may alternatively suggest the possibility
of an unusually low-mass white dwarf.  If we assume that the
secondary of J1944 has a normal mass for this orbital period,
such as 0.066$M_\odot$ in WZ Sge \citep{kat13qfromstageA},
the mass of the white dwarf must be $\sim$0.47$M_\odot$.
According to \citet{zor11SDSSCVWDmass}, the fraction of CVs
having white dwarf lighter than 0.5$M_\odot$ is only
7$\pm$3 \%, even including suspicions measurements.
Furthermore, there is evidence from modern eclipse observations
that mass of the white dwarf in short-$P_{\rm orb}$ CVs 
is not diverse \citep{sav11CVeclmass}.
We therefore consider the interpretation of a massive secondary
more likely.

   The presence of precursor outburst and the high frequency
of normal outbursts are usual features of longer-$P_{\rm orb}$
systems such as V1504 Cyg and V344 Lyr.  Systems with
$P_{\rm orb}$ like J1944 are usually WZ Sge-type dwarf novae
with very rare (super)outbursts (e.g. \cite{kat01hvvir})
or ER UMa-type dwarf novae, a rare subgroup with very frequent
outbursts and short supercycles (e.g. \cite{kat95eruma};
\cite{rob95eruma}).  J1944 does not match the properties
of either group.  This can be understood if the outburst
properties are a reflection of the mass ratio rather than
the orbital period since the $q$ value of J1944 is closer
to those of longer-$P_{\rm orb}$ SU UMa-type dwarf novae.

   The presence of such a system would pose a problem
in terms of the CV evolution since the secondary loses
its mass during the CV evolution and $q$ value is expected
to be as low as $\sim$0.08 around the orbital period
of J1944.  In recent years, some objects showing hydrogen lines
in their spectra (this excludes the possibility of
double-degenerate AM CVn-type systems)
have been discovered around this period or even in shorter
period.  These objects include EI Psc (\cite{uem02j2329letter};
\cite{tho02j2329}), V485 Cen \citep{aug96v485cen}
and GZ Cet \citep{ima06j0137}.  These objects are considered
to be CVs whose secondary had an evolved core at the time
of the contact, and are considered to be progenitors
of AM CVn-type double white dwarfs (\cite{pod03amcvn};
\cite{nel04amcvn}; \cite{uem02j2329letter}; \cite{tho02j2329}).

   None of these objects have been reported for $q$ determination
directly from radial-velocity studies, and $q$ values have
only been inferred from the traditional $\varepsilon$,
which has an unknown uncertainty \citep{kat13qfromstageA}.
The detection of stage A superhumps in J1944 allowed
the first reliable determination of $q$ in such stripped-core 
ultracompact binaries.  There is, however, a marked difference
of the outburst frequency between J1944 and these known objects
since the frequency of outbursts in such systems have been
reported to be low \citep{tho13j1340}.
This suggests that J1944 has an anomalously high mass-transfer
rate among these objects.  The object may be
in a phase analogous to ER UMa-type dwarf novae, whose
high mass-transfer rates may be a result of a recent 
classical nova explosion 
(cf. \cite{kat95eruma}; \cite{pat13bklyn}).
Since the object can be within
the reach of the ground-based telescopes, the exact optical
identification and the search for the feature of
the secondary star are encouraged to solve the mystery.

\medskip

We thank the Kepler Mission team and the data calibration engineers for
making Kepler data available to the public.
We also thank the Planet Hunters group for making their information
on the background dwarf novae public which enabled us to 
study this interesting object. 
This work was supported by the Grant-in-Aid
``Initiative for High-Dimensional Data-Driven Science through Deepening
of Sparse Modeling'' from the Ministry of Education, Culture, Sports, 
Science and Technology (MEXT) of Japan.


\begin{thebibliography}{}

\bibitem[Augusteijn et~al.(1996)]{aug96v485cen}
  Augusteijn, T., van~der Hooft, F., de Jong, J.~A., \& van Paradijs, J.\ 1996,
  A\&A, 311, 889

\bibitem[{Barclay} et~al.(2012)]{bar12j1939}
  {Barclay}, T., {Still}, M., {Jenkins}, J.~M., {Howell}, S.~B., \&
  {Roettenbacher}, R.~M.\ 2012, MNRAS, 422, 1219

\bibitem[{Borucki} et~al.(2010)]{bor10Keplerfirst}
  {Borucki}, W.~J., {et~al.}\ 2010, Science, 327, 977

\bibitem[{Cleveland}(1979)]{LOWESS}
  {Cleveland}, W.~S.\ 1979, J. Amer. Statist. Assoc., 74, 829

\bibitem[{Fischer} et~al.(2012)]{fis12PlanetHunters}
  {Fischer}, D.~A., {et~al.}\ 2012, MNRAS, 419, 2900

\bibitem[{Imada} et~al.(2006)]{ima06j0137}
  {Imada}, A., {et~al.}\ 2006, PASJ, 58, 143

\bibitem[{Kato} et~al.(2009)]{Pdot}
  {Kato}, T., {et~al.}\ 2009, PASJ, 61, S395

\bibitem[{Kato}, {Kunjaya}(1995)]{kat95eruma}
  {Kato}, T., \& {Kunjaya}, C.\ 1995, PASJ, 47, 163

\bibitem[Kato et~al.(1999)]{kat99erumareview}
  Kato, T., Nogami, D., Baba, H., Masuda, S., Matsumoto, K., \& Kunjaya, C.\
  1999, in Disk Instabilities in Close Binary Systems, ed. S. Mineshige, \&
  J.~C. Wheeler (Tokyo: Universal Academy Press), p.~45

\bibitem[{Kato}, {Osaki}(2013a)]{kat13j1939v585lyrv516lyr}
  {Kato}, T., \& {Osaki}, Y.\ 2013a, PASJ, 65, 97

\bibitem[{Kato}, {Osaki}(2013b)]{kat13qfromstageA}
  {Kato}, T., \& {Osaki}, Y.\ 2013b, PASJ, 65, 115

\bibitem[{Kato} et~al.(2001)]{kat01hvvir}
  {Kato}, T., {Sekine}, Y., \& {Hirata}, R.\ 2001, PASJ, 53, 1191

\bibitem[{Knigge} et~al.(2011)]{kni11CVdonor}
  {Knigge}, C., {Baraffe}, I., \& {Patterson}, J.\ 2011, ApJS, 194, 28

\bibitem[{Koch} et~al.(2010)]{Kepler}
  {Koch}, D.~G., {et~al.}\ 2010, ApJ, 713, L79

\bibitem[{Martin} et~al.(2005)]{GALEX}
  {Martin}, D.~C., {et~al.}\ 2005, ApJ, 619, L1

\bibitem[{Nelemans} et~al.(2004)]{nel04amcvn}
  {Nelemans}, G., {Yungelson}, L.~R., \& {Portegies Zwart}, S.~F.\ 2004, MNRAS,
  349, 181

\bibitem[{Osaki}(1989)]{osa89suuma}
  {Osaki}, Y.\ 1989, PASJ, 41, 1005

\bibitem[{Osaki}, {Kato}(2013)]{osa13v344lyrv1504cyg}
  {Osaki}, Y., \& {Kato}, T.\ 2013, PASJ, 65, 95

\bibitem[{Osaki}, {Kato}(2014)]{osa14v1504cygv344lyrpaper3}
  {Osaki}, Y., \& {Kato}, T.\ 2014, PASJ, 66, 15

\bibitem[{Patterson} et~al.(2013)]{pat13bklyn}
  {Patterson}, J., {et~al.}\ 2013, MNRAS, 434, 1902

\bibitem[Podsiadlowski et~al.(2003)]{pod03amcvn}
  Podsiadlowski, Ph., Han, Z., \& Rappaport, S.\ 2003, MNRAS, 340, 1214

\bibitem[Robertson et~al.(1995)]{rob95eruma}
  Robertson, J.~W., Honeycutt, R.~K., \& Turner, G.~W.\ 1995, PASP, 107, 443

\bibitem[{Savoury} et~al.(2011)]{sav11CVeclmass}
  {Savoury}, C.~D.~J., {et~al.}\ 2011, MNRAS, 415, 2025

\bibitem[{Thorstensen}(2013)]{tho13j1340}
  {Thorstensen}, J.~R.\ 2013, PASP, 125, 506

\bibitem[Thorstensen et~al.(2002)]{tho02j2329}
  Thorstensen, J.~R., Fenton, W.~H., Patterson, J.~O., Kemp, J., Krajci, T., \&
  Baraffe, I.\ 2002, ApJ, 567, L49

\bibitem[Uemura et~al.(2002)]{uem02j2329letter}
  Uemura, M., {et~al.}\ 2002, PASJ, 54, L15

\bibitem[{Zorotovic} et~al.(2011)]{zor11SDSSCVWDmass}
  {Zorotovic}, M., {Schreiber}, M.~R., \& {G{\"a}nsicke}, B.~T.\ 2011, A\&A,
  536, A42

\end{thebibliography}
\end{document}